\documentclass{appolb}
\usepackage{graphicx}
\usepackage[detect-all]{siunitx}

\begin{document}
\title{Diamond detectors
	for radiation monitoring and beam abort
	at Belle II%
\thanks{presented at XXVI Cracow EPIPHANY Conference on LHC Physics: SM and Beyond}%
}
\author{Riccardo Manfredi
\address{University and INFN Trieste
\\
On behalf of the Belle II collaboration
}
}
\maketitle
\begin{abstract}
\noindent The Belle II experiment will be at the forefront of indirect searches for non-Standard-Model physics using billions of heavy quarks and $\tau$ leptons produced in high-intensity 10 GeV electron-positron collisions from the SuperKEKB collider.
The intense beams needed to achieve the required precisions are associated with high beam-background radiation that may damage the inner detectors. 
A dedicated radiation-monitoring and beam-abort system, based on artificial diamond sensors, ensures protection and safe data taking conditions. 
I briefly outline the system and illustrate the operational experience and performance during 2019 physics operations.
\end{abstract}

\section{Introduction}
Belle II is an international collaboration that aims to indirectly test the Standard Model by precisely studying billions of decays of $\tau$ leptons and mesons containing $b$ and $c$ quarks produced in high-energy collisions.\\
The experiment is located at the KEK laboratory in Tsukuba, Japan, on the SuperKEKB accelerator, a high-luminosity $e^+e^-$ energy-asymmetric collider \cite{acc_design}, designed to produce nearly 900 $B\overline{B}$ pairs ($B$ = $B^0$ or $B^+$) per second via decays of $\Upsilon(4S)$ mesons produced at threshold. The experiment plans to collect $\SI{50}{ab^{-1}}$ of integrated luminosity by 2029, corresponding to $\num{5e10}$ $B\overline{B}$ pairs. The reduction of the luminous-volume size by a factor of 20 at SuperKEKB with respect to its predecessor KEKB, combined with a doubling of beam currents, is expected to yield a gain in luminosity by a factor of 40. This is achieved with a nano-beam collision scheme based on small horizontal and vertical emittance, as well as a large crossing angle \cite{nanobeam}.

\medskip
The Belle II detector, an upgrade of its predecessor Belle, has a cylindrical geometry. The inner subdetectors are dedicated to the reconstruction of charged-particle trajectories (tracks) and the outer subdetectors are used for neutral-particle reconstruction and charged-particle identification. 
Efficient and precise track reconstruction is a key performance driver for Belle II, since charged particles are common in flavor-physics final states and analyses rely strongly on precise measurements of their momenta and the decay positions of their long-lived parent particles.\\
The 2018 pilot-run, with an incomplete detector, validated the collision scheme. The 2019 operations aimed at taking physics data with the full detector, while steadily increasing the instantaneous luminosity. By the end of 2019 luminosity reached $\SI{1.14 e34}{\centi\meter^{-2}\second^{-1}}$, about half of Belle's record peak luminosity \cite{Bfactories}.\\
The high luminosity required to reach Belle II goals has the drawback of intense radiation from beam losses. At present, the dominant processes contributing to these losses are interactions of particles with each other inside each bunch or with residual gas atoms in the beam pipe, or the emission of synchrotron radiation \cite{Fondi}. Such ``beam background'' illuminates the active parts of the detector, hence it needs to be monitored and mitigated to protect the detector and the accelerator.

\section{Diamond system for radiation monitoring and beam abort}
The vertex detector can tolerate a maximum of $10 - 20$ Mrad in a decade of operations, but high radiation spikes ($\SI{1}{rad}$ or more in less than 1 ms) can induce localized damage. Hence, a reliable protection system is key to keep it safe from beam-background radiation and ensure its optimal performance for the expected lifetime. In addition, such a system provides protection to the SuperKEKB final-focus superconducting magnets, which heat up and lose their focusing properties (quench) if hit by significant background.

\begin{figure}[htb]
\centerline{%
\includegraphics[width=0.9\textwidth]{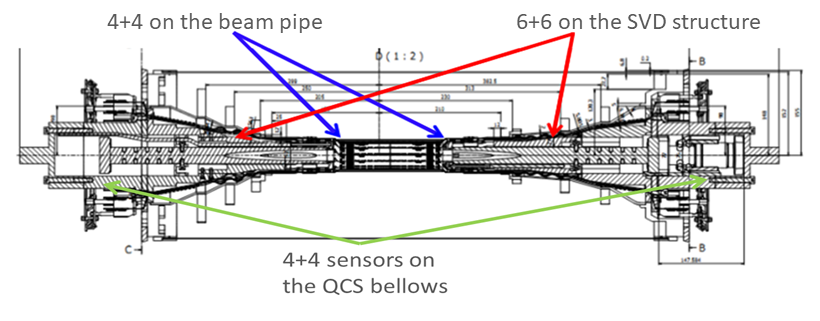}}
\caption{\footnotesize{Belle II diamond sensor configuration in 2019. Eight sensors are installed on the beam pipe, twelve on the silicon tracker support, and eight on the focusing magnets bellows.}}
\label{fig:disposizione}
\end{figure}

\medskip
The Belle II radiation-monitoring and beam-abort system is based on synthetic single-crystal diamond sensors. Each sensor works as a solid-state drift chamber: traversing charged particles free electron-hole pairs that drift due to an applied bias voltage of $\mathcal{O}(100) \si{V}$ and are collected on the metallic ends of the sensors \cite{berdermann2014diamond}. The resulting time-dependent current is amplified by a trans-impedance amplifier and digitized.\\
Many properties make diamonds suitable for this purpose: (i) radiation resistance, which ensures reliable and stable long-term operations; (ii) a rapid response time, which allows for detecting sudden intense radiation bursts; (iii) a broad range of currents (pA to mA), corresponding to dose rates from $\si{\mu rad/s}$ to $\mathcal{O}(10)$ krad/s; and (iv) small size, to fit the tight space constraint.\\
The system is composed of 28 sensors, installed in various positions around the beam pipe, as shown in Fig \ref{fig:disposizione}.\\

\begin{figure}[htb]
	\centerline{%
		\includegraphics[width=0.63\textwidth]{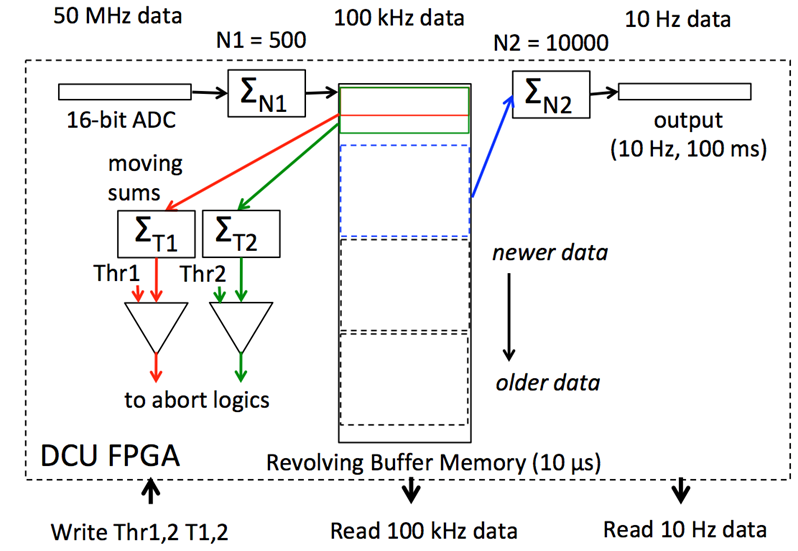}}
	\caption{\footnotesize{Block diagram representing the FPGA data stream.}}
	\label{fig:FPGA}
\end{figure}

\medskip
Each group of four diamonds is controlled and read out by an FPGA, which controls the dynamic range of the amplifier, as well as providing digitizer section and provides data for radiation monitoring (10 Hz) and the beam-abort logic (100 kHz). Figure \ref{fig:FPGA} illustrates the FPGA logic. The amplified diamond currents, digitized and oversampled at 50 MHz by ADCs, are summed in blocks of 500 samplings to achieve the 100 kHz-sampling data stream used in the abort logic. Two different moving sums run on this data stream, adding the latest value and discarding the oldest one every $\SI{10}{\mu s}$: these sums achieve an integration over moving time windows or ``gates''. If one of the resulting sums exceeds the corresponding abort threshold, a beam-abort request is sent. Both threshold values and integrating gate lengths are configurable. In addition, the 100 kHz data are summed in blocks of $10^4$, resulting in a 10 Hz data-stream that is saved and is used for online and offline radiation monitoring.\\
In parallel, other independent systems, controlled by the accelerator, can issue beam aborts. Whenever a beam abort is issued, the FPGA memory pointer freezes and the last second of data, sampled at 100 kHz, is written and saved for post-abort studies. 

\section{Operations}
\begin{figure}[htb]
	\begin{minipage}{0.5\textwidth}
		\centering
		\includegraphics[width=1.1\linewidth]{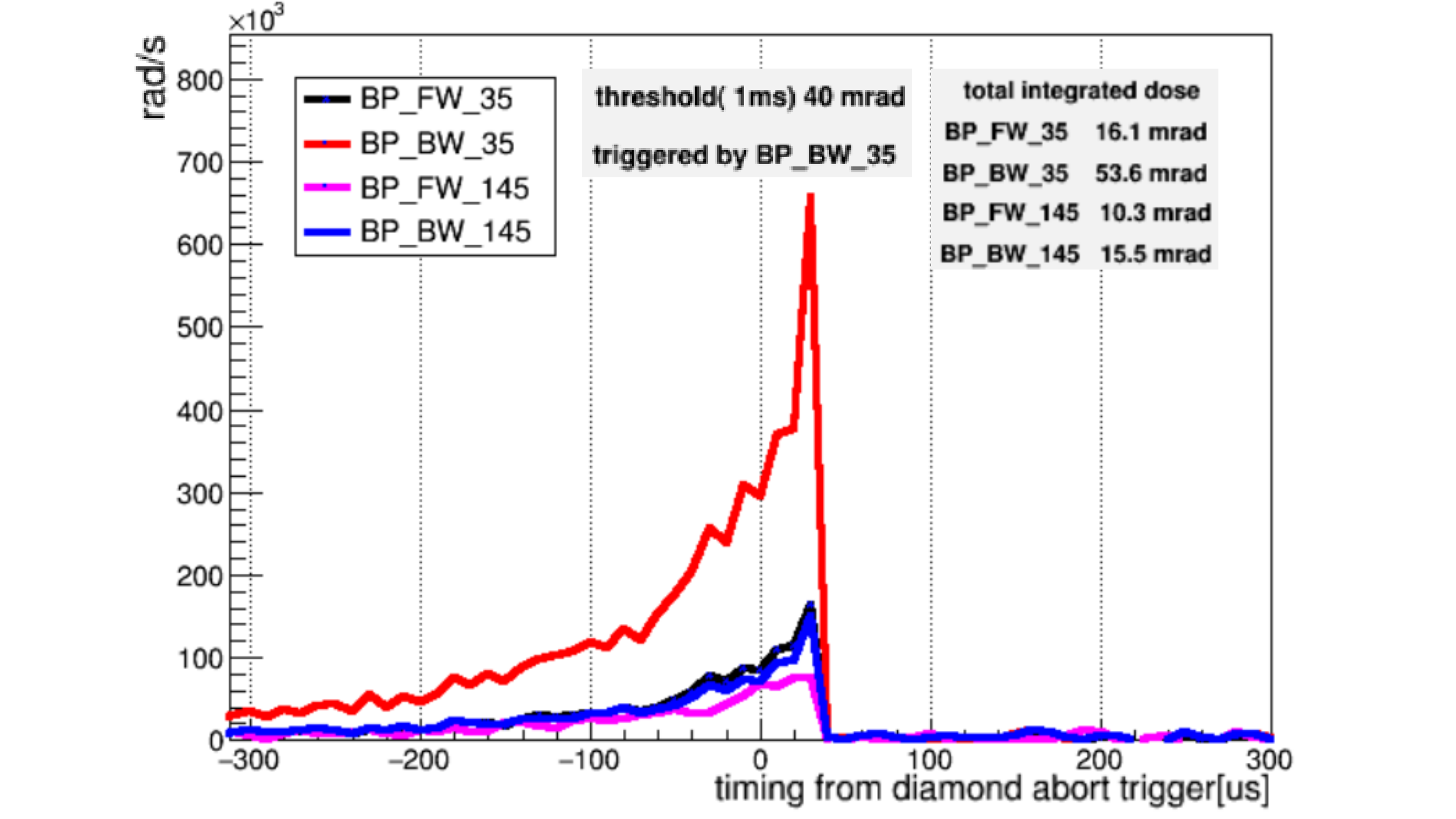}
	\end{minipage}
	\begin{minipage}{0.5\textwidth}
		\centering
		\includegraphics[width=1.15\linewidth]{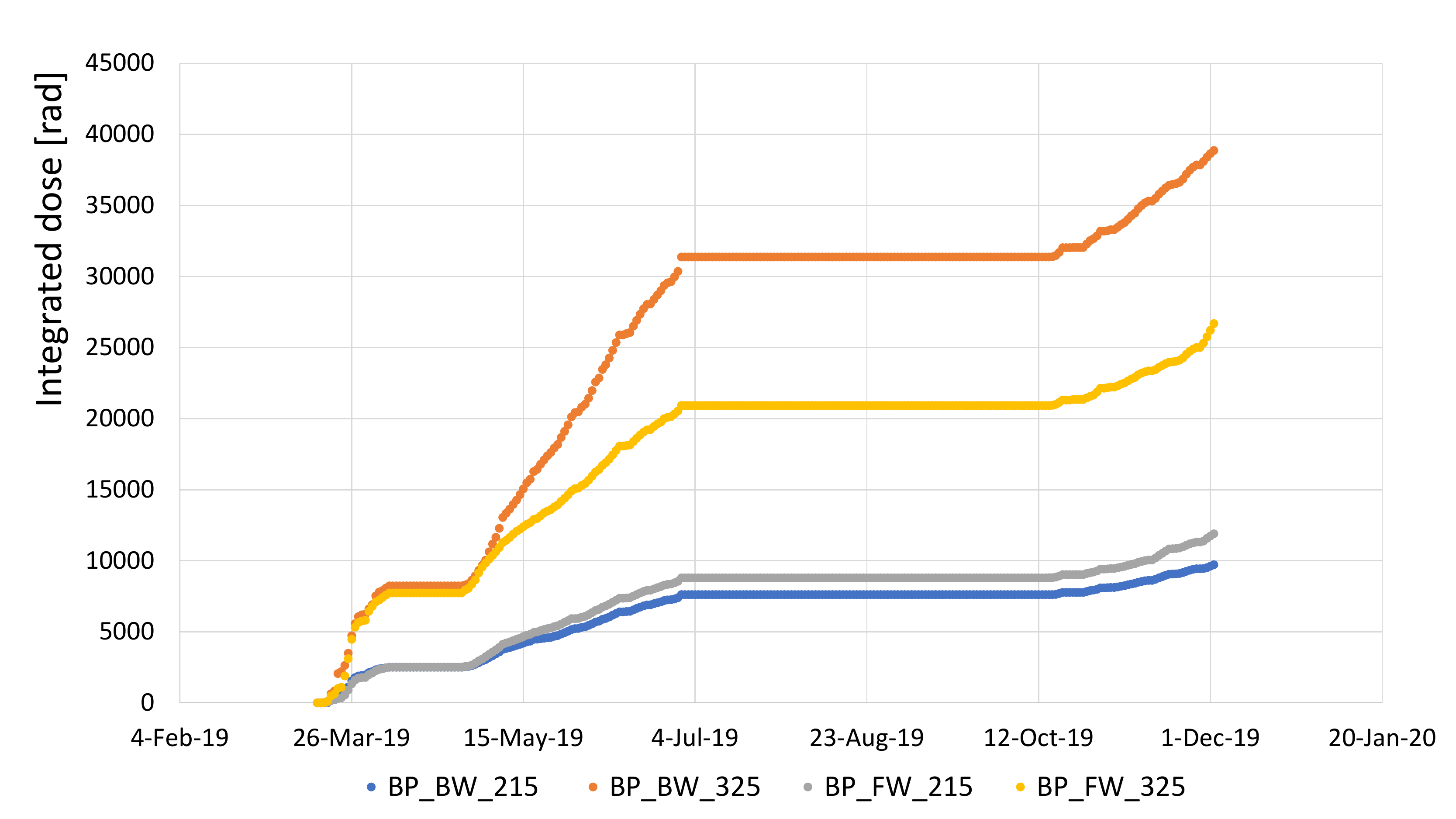}
	\end{minipage}
	\caption[caption]{\footnotesize{(left)  Dose rate as a function of time seen by four diamonds installed on the beam pipe. They correspond to a diamond beam-abort triggering event. (right) Total integrated dose as a function of time.}}
	\label{fig:ops}
\end{figure}

The goal is to achieve the optimal trade-off between the safety of the detector and efficient data-taking: beam-aborts should be issued whenever beam-background losses are too high, but triggering a beam-abort when conditions are safe (namely, triggering ``fake aborts'') must be avoided. Equilibrium is achieved through continuous monitoring of diamond performances and consistent setting of pedestals, beam-abort thresholds, and integrating gate lengths, while adapting to the continuous variation of accelerator parameters.\\
The first main task is \textbf{beam aborting}; an example of diamonds dose rates triggering a beam abort is shown in the left plot of Fig. \ref{fig:ops}. When a beam-abort signal is generated from the diamonds installed at the interaction point, it is sent via $\SI{700}{\meter}$ fibre-optic cable to the SuperKEKB control room and logically combined with signals coming from all other accelerator radiation monitors. Then an activation signal is sent to the kicker magnets, which dump the beams. The diamond system issued about 300 beam aborts in 6 months of 2019 operations: these events are not uniformly distributed in time, but are concentrated during periods of machine study that stress the accelerator conditions beyond those used for physics running.\\
The other important task of the system is \textbf{radiation monitoring.} The 10 Hz data stream is used for online and offline monitoring of correlations between radiation doses read in diamonds and other accelerator variables, such as beam currents or collimator openings. Saved data are also used to evaluate the integrated doses over time, which are used to extrapolate the doses in the inner layers of the detector. The  total integrated dose in 2019 is shown as a function of time in the right plot of Fig. \ref{fig:ops}.\\

\begin{figure}[htb]
	\centerline{%
		\includegraphics[width=0.8\textwidth]{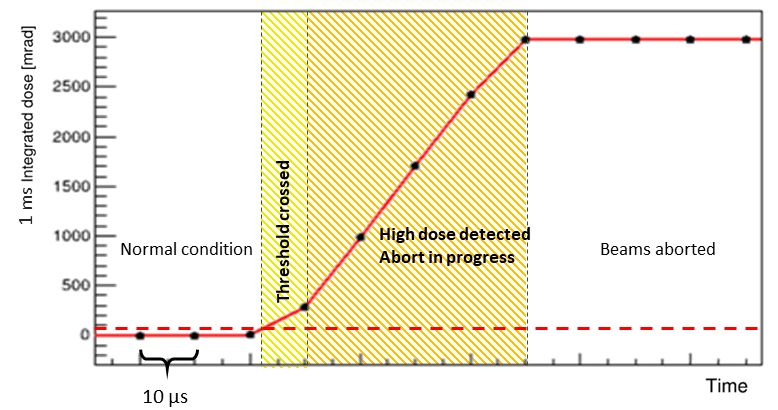}}
	\caption{\footnotesize{Integrated dose as a function of time for one sensor. The dose is integrated over a moving 1 ms-gate, updated every $\SI{10}{\micro \second}$. The dashed red line indicates the threshold.}}
	\label{fig:severe_loss}
\end{figure}

\medskip
To further optimize the system and improve protection, we studied the phenomenology of some extraordinary events too. An example is a severe beam loss that occurred in June 2019, probably due to a beam-dust scattering event; Figure \ref{fig:severe_loss} shows the integrated dose in the integration gate of the abort logic as a function of time. Most of the dose has been integrated after exceeding the threshold. Two delays are visible:  the time between the threshold passing and the abort-request signal generation (yellow hatching in Fig.\ref{fig:severe_loss}), and the time between the beam-abort trigger from diamonds and the completion of the beam dump by the abort kicker magnets (orange hatching in Fig.\ref{fig:severe_loss}). These studies prompted the need for improving timing both in the diamond and accelerator abort logic \cite{Manfredi:2019lna}. 

\medskip
During the 2019 winter shutdown, the FPGA firmware was upgraded to reduce such delays in the beam-abort chain. The abort-logic data-stream changed from 100 kHz to 400 kHz sampling, i.e., the dose rates are now compared with the thresholds every $\SI{2.5}{\micro\second}$.\\
This upgrade led to multiple improvements of the diamond system: the delay between the instants a threshold is crossed and diamonds trigger an abort is reduced, and a shorter time integration for the abort logic is now available. In addition, a four-fold improvement is gained in the time resolution of post-abort memory data, at the cost of an acceptable noise increase.

\section{Summary}
The Belle II experiment will improve our understanding of flavor physics during the next decade, which will guide indirect searches for the completion of the Standard Model at the intensity frontier. Its success critically depends on controlling the beam backgrounds and protecting the detector from them.\\
A system based on 28 diamond sensors, installed around the interaction region, provides excellent radiation-monitoring and beam-abort capabilities. The system performed well during the 2019 spring and fall runs, and has been improved in preparation of the 2020 data taking. It now features a finer dose sampling, reduced delays in the abort chain and an overall optimized configuration. The diamond system will ensure Belle II collects data safely and efficiently, while offering useful feedbacks to the accelerator. 


\bibliographystyle{belle2-bibliography}
\bibliography{biblio}

\end{document}